PCP-15, for *PHYSICA C*

# Anomalous lattice expansion of $RuSr_2Eu_{1.5}Ce_{0.5}Cu_2O_{10-\delta}$ (Ru-1222) magneto superconductor: A low temperature X-ray diffraction study

V.P.S. Awana[1,2,*], Rajeev Ranjan[3], Rajeev Rawat[4], L. S. Sharath Chandra[4], M. Peurla[5], V. Ganesan[4], H. Kishan[2], D. Pandey[3], R. Laiho[5], E. Takayama-Muromachi[1], and A.V. Narlikar[4,5]

[1]Superconducting Materials Center, NIMS, 1-1 Namiki, Tsukuba, Ibaraki, 305-0044, Japan

[2]National Physical Laboratory, K.S. Krishnan Marg, New Delhi-110012, India

[3]School of Materials Science and Technology, Institute of Technology, Banaras Hindu University, Varanasi-221005, India

[4]UGC-DAE Consortium for Scientific Research, University Campus, Khandwa Road, Indore-452017, MP, India

[5]Wihuri Physical Laboratory, University of Turku, FIN - 20014, TURKU, Finland

Detailed X-ray diffraction studies on $RuSr_2Eu_{1.5}Ce_{0.5}Cu_2O_{10-\delta}$ (Eu-Ru-1222) magneto superconductor down to 13 K at close temperature intervals revealed that with decreasing temperature, the c-parameter decreases linearly down to 100 K, following the relation $c = 28.412(1) + 2.5(1)\text{x}10^{-4}T$ and then becomes almost temperature independent on further lowering of temperature. Interestingly, the linear decrease in the *a*-parameter is seen only up to 180 K. Below 180 K, the observed values of the *a*-parameters are more than the value expected from the linear variation. This departure can be considered as development of strain in the system. Interestingly enough, these temperatures coincide with the complex magnetic ordering ($T_{mag}$) of Ru in Ru-1222 system. This is the first report of the observation of the onset of excess volume and also of the strain along the *a*-axis near the magnetic ordering temperature in Ru-1222 superconductor, *and indicates a* coupling between the lattice and the magnetism in this system. Magnetization, magneto transport and thermoelectric power measurements being carried out on the same sample are also reported. The results of low temperature excess volume and the strain along *a* –axis of Ru-1222 compound are reportedly not seen in Ru-1212 ($RuSr_2GdCu_2O_8$).

* Corresponding Author: e-mail; awana@mail.nplindia.ernet.in & AWANA.veer@nims.go.jp

**INTRODUCTION**

Rutheno-cuprates are a topic of discussion for over half a decade by now due to their co-existing superconducting and magnetic properties [1, 2]. Both Ru-1222 and Ru-1212 phases are structurally related to the popularly known Y:123 ($YBa_2Cu_3O_{7-\delta}$) or Cu-1212, e.g. $CuBa_2YCu_2O_{7-\delta}$ phase with Cu in the charge reservoir replaced by Ru such that the Cu-O chain is replaced by a $RuO_2$ sheet. In the Ru-1222 structure furthermore, a three-layer fluorite-type block instead of a single oxygen-free *R* (= rare earth element) layer is inserted between the two $CuO_2$ planes of the Cu-1212 structure [3-5]. In both Ru-1222 and Ru-1212 the magnetism is originated basically from $RuO_2$ sheet or more precisely the $RuO_6$ octahedras and superconductivity is supposed to reside in $CuO_2$ planes. Ferromagnetically (canted) ordered Ru spins at 5 K imposes an internal field over the whole unit cell [2] and brings the compound into spontaneous vortex phase (SVP) even in zero external field [6]. This has been demonstrated experimentally in Ru-1222 by optical magnetometry [7]. The key questions yet to be addressed in these systems are: (a) the genuine co-existence of two order parameters (superconducting and magnetic), (b) the nature of magnetism of Ru spins.

As far as genuine co-existence is concerned few doubts are still cast related to phase separation [3, 8, 9] of the two phenomenon. However majority of the data from scientific community indicates towards genuine co-existence [1, 2, 6, 7, 10-13]. At the same time, presently there seems no clear picture of magnetism of Ru spins in these compounds (Ru-1212/1222), and interestingly the most recent magnetization results have caste some doubts on earlier studies related to magnetism of Ru-1212 [13-15]. Also the so-called anti-ferro/ferromagnetic transition at around 140 K is not seen thermodynamically in heat capacity ($C_P$) measurements which reveal only a broad hump up to room temperature spread over a range of around 150 K [15]. As far as Ru-1222 is concerned, the main features are the same as for both Ru-1212. The magnetic structure of Ru-1222 has been studied by neutron powder diffraction [16]. Despite the fact that host of physical-property measurements have been carried out on Ru-1212 [2, 4, 8-15] and Ru-1222 [1, 6, 7, 16-19], no final consensus so far seems to be in sight. One of the important parameters affecting the transport properties (electrical, thermal, magnetic etc.) is the variation of lattice parameters of the compound. To understand their temperature dependence its structure including lattice constants need to be probed simultaneously in the pertinent low temperature range. In this connection, it is worth mentioning that the Ru-1212 compound when probed using neutron diffraction down to 35 K, surprisingly revealed no change in the lattice constants in relation to the magnetic ordering temperature ($T_{mag}$) of Ru. The variation (decrease) of lattice parameters was smoothened down to 35 K [20].

Though the magnetic structure of Ru-spins in Ru-1212 and Ru-1222 seems similar, there are many finer differences**.** For instance the spin-glass (SG) component is not seen in the former but is present in later [21]. More interestingly the lattice parameters did not respond to $T_{mag}$ of Ru in Ru-1212 [20]. Here in the present article we report the variation of lattice parameters with temperature down to 13 K for Eu-Ru-1222 system. We observed an onset of excess volume and strain along the *a*-axis near the magnetic ordering temperature of Ru-1222 superconductor. *This indicates a possible* coupling between the lattice and the magnetism in this system. This interesting observation on Ru-1222 may help in explaining the complicated and yet not well understood magnetism of Ru in Ru-1222 compounds.

## EXPERIMENTAL DETAILS

The $RuSr_2Eu_{1.5}Ce_{0.5}Cu_2O_{10-\delta}$ (Eu-Ru-1222) sample was synthesized through a solid-state reaction route from $RuO_2$, $SrO_2$, $Eu_2O_3$, $CeO_2$ and CuO. Calcinations were carried out on the mixed powder at 1000, 1020, 1040 and 1060 $^0$C each for 24 hours with intermediate grindings. The pressed bar-shaped pellets were annealed in a flow of oxygen at 1075 $^0$C for 40 hours and subsequently cooled slowly over a span of another 20 hours down to room temperature. X-ray diffraction (XRD) patterns were obtained using $CuK_\alpha$ radiation down to 10 K with close temperature intervals. Magnetization measurements were performed on a SQUID magnetometer (Cryogenic Ltd. model S600). Resistivity measurements under applied magnetic fields of 0 to 6 Tesla were made in the temperature range of 5 to 300 K using a four-point-probe technique. Thermoelectric power (TEP) measurements were carried out by dc differential technique over a temperature range of 5 – 300 K, using a home made set up. Temperature gradient of ~1 K was maintained throughout the TEP measurements.

## RESULTS AND DISCUSSION

The magnetic susceptibility $\chi$ vs. temperature $T$ behaviour in the temperature range of 5 to 300 K for the Eu-Ru-1222 sample under applied field of 10 Oe, measured in both zero-field-cooled (ZFC) and field-cooled (FC) modes for Eu-Ru-1222 compound is exhibited in Fig.1. The ZFC and FC curves start branching at around 150 K with a sharp upward turn around 100 K. This means that the magnetic transition temperature of Eu-Ru-1222 is $T_{mag}$ = 150 K. The ZFC branch shows further a cusp at $T_{FM/SG}$ = 80 K, then the superconducting transition temperature at $T_{c,\chi}$ = 25 K and finally a diamagnetic transition at $T_d$ = 20 K. $T_{c,\chi}$ is seen as a step in FC transition, for more detailed description, please see refs. [1, 3, 18, 19, 22]. Generally speaking the $\chi$ ($T$) behaviour of presently studied compound is similar to that as reported earlier [1, 7, 18, 19]. The isothermal magnetization as a function of magnetic field at 5 K may be viewed as: $M(H) = \chi H + \sigma_s(H)$, where $\chi H$ is the linear contribution from antiferromagnetic ($T_{mag}$) Ru spins and also from paramagnetic Eu spins, and $\sigma_s(H)$ represents the ferromagnetic component ($T^{FM/SG}$) of Ru. The appearance of the ferromagnetic component at low temperatures within antiferromagnetic/ spin-glass Ru spins could happen due to a slight canting of the spins, as seen from neutron diffraction for Ru-1212 [10]. The overall magnetic behaviour of Eu-Ru-1222 has been explained earlier by some of us and also others [1, 3, 7, 18, 19, 22]. In short, the Ru spins order antiferro- magnetically (AFM) at around 150 K with canted ferromagnetism (FM) below $T_{cusp}$ = 85 K. Near $T_{cusp}$ the compound is known to exhibit the spin-glass (SG) behaviour due to competing AFM and FM components [19].

Resistivity versus temperature (ρ($T$)) plot with and without magnetic field for the pristine Eu-Ru-1222 sample is shown in upper inset of figure 1. This sample exhibits a slight semiconducting behaviour down to 200 K and below that it is metallic with a small positive slope (extended plot not shown, for details please see ref. 23). Interestingly the ρ($T$) behavior is nearly constant just above the superconducting transition temperature onset ($T_c^{\text{onset}}$). The compound shows $T_c^{\text{onset}}$ of around say 33 K and $T_c^{\rho=0}$ at 25 K. The general ρ($T$) behaviour is reminiscent of an underdoped high $T_c$ cuprate system and is in general agreement with other reports on $RuSr_2Eu_{1.5}Ce_{0.5}Cu_2O_{10-\delta}$ [1, 7, 18, 19]. Under applied magnetic field though the $T_c^{\text{onset}}$ remains nearly unchanged the $T_c^{\rho=0}$ value goes down to 5 K under applied magnetic field of 8 T. The rate of $T_c$ decreament is nearly 2.5 K/T. Estimated values of mean free path ($l$), upper critical field ($H_{c2}$ (0)) and Ginzburg-Landau coherence length ($\xi(0)$) are 56 Å, 55 T and 24.5 Å respectively, for details please see ref. 23. This means that the present sample of $RuSr_2Eu_{1.5}Ce_{0.5}Cu_2O_{10-\delta}$ is in the clean limit. Thermoelectric power versus temperature $S(T)$ plot of the compound is shown in Fig. 2. The room temperature value of $S$ i.e. $S^{300K}$ is found to be around 24 μV/K. $S$ passes through a maximum at around 170 K, being denoted in figure as TEP peak temperature. Detailed analysis of $S(T)$ is provided in ref. 23.

The powder x-ray diffraction patterns of the $RuSr_2Eu_{1.5}Ce_{0.5}Cu_2O_{10-\delta}$ were recorded at various temperatures in the range 13 K to 293 K, plot not shown. The room temperature powder diffraction pattern of this compound resembles that of other isostructural compounds reported in the literature [24-26]. The entire pattern can be indexed on the basis of a body centered tetragonal cell with space group I4/*mmm*. Felner et al [26] reported similar results on $RE_{1.5}Ce_{0.5}RuSr_2Cu_2O_{10}$ (RE= Eu, Gd). Rietveld refinement of the structure was carried out with the model used by Tamura et al [24]. The program Fullprof [27] was used for the structure refinement. Pseudo-voigt profile function was chosen to model the individual Bragg profiles. Background was fitted with a sixth order polynomial function. The isotropic thermal parameters of oxygen were fixed to be 1.0Å$^2$.

The refinements at other temperatures were carried out in a sequential manner, i.e., the final refined structural parameters obtained at a particular temperature was used as initial input parameters for the next lower temperature and so on. Fig. 3 depicts the variation of the lattice parameters obtained using the Rietveld analysis of the XRD data. As is evident from this figure, with decreasing temperature, the *c*-parameter decreases linearly down to 100 K, following the relation $c = 28.412(1) + 2.5(1)\text{x}10^{-4}T$ and then becomes almost temperature independent on further lowering of temperature. Interestingly, the linear decrease in the *a*-parameter is seen only upto 180 K. Below 180 K, the observed values of the *a*-parameters are more than the value expected from the linear variation. This departure can be considered as development of strain in the system. A plot of this strain, defined as $(\Delta a/a)$x100, with temperature is shown in the inset to Fig. 3. The onset temperature is somewhere between 180 K and 160 K which is close to the first magnetic ordering temperature ($T_{mag}$) (please see Fig. 1, ZFC and FC branching temperature). The steep rise in the strain below 100 K is due to the fact that, similar to *c*-parameter, the *a*-parameter has become almost temperature independent. Identical trend is also observed in the temperature variation of the unit cell volume (see Fig. 4). The excess volume ($\Delta V$), calculated as the difference between the observed value and the value predicted from the extrapolation of the linear variation (see Fig.4), exhibits a temperature variation identical to percentage *a*-strain. Further details of the lattice parameter variations and ensuing strain etc. will be published in a more detailed article [28].

It is clear that *a*-parameter is linear only down to $T_{mag}$, which is the onset of the AFM ordering of Ru spins, whether *c*-parameter is linear down to 100 K, corresponding to the evolution of canted FM component in the compound. Another interesting fact is that though non-linearity in *a*-parameter starts below say 150 K, its sharp deviation is seen only below 100 K. This has given rise to steep change in the strain, defined as ($\Delta a/a$)x100. It seems at $T_{mag}$, small deviation of only *a*-parameter and not the *c* is due to the indication of 2D short range AFM ordering, which is in agreement with Mossbauer spectroscopy results [29]. Both *a* and *c*-parameters deviate sharply from their linear decrease with temperature below 100 K. This temperature corresponds to $T_{FM/SG}$. It means the 3D steep rise is seen in strain/volume below the setting of canted FM component within the AFM structure.

## SUMMARY

In summary, detailed X-ray diffraction study on $RuSr_2Eu_{1.5}Ce_{0.5}Cu_2O_{10-\delta}$ (Eu-Ru-1222) magneto superconductor down to 13 K, *indicates* the observation of an onset of the excess volume and also of the strain along the a-axis near the magnetic ordering temperature in Ru-1222 superconductor. *There is possibility* that the lattice and the magnetism are coupled in this system.

## ACKNOWLEDGEMENT

This work is partially supported by INSA-JSPS bilateral exchange visit of Dr. V. P. S. Awana to NIMS Japan. One of us (AVN) thanks Prof. Laiho and the University of Turku for providing research facilities and for the invitation for the present visit.

**FIGURE CAPTIONS**

Fig. 1. $\chi$ *(T)* behaviour for the $RuSr_2Eu_{1.5}Ce_{0.5}Cu_2O_{10-\delta}$. The inset shows the resistivity $\rho$ (*T*) behaviour of the same compound in various applied fields.

Fig. 2 *S* (*T)* behaviour for the $RuSr_2Eu_{1.5}Ce_{0.5}Cu_2O_{10-\delta}$.

Fig. 3. Variation of *a* (solid circles) and *c* (open circles) parameters of $RuSr_2Eu_{1.5}Ce_{0.5}Cu_2O_{10}$ with temperature. The straight lines are fits to the data points. The respective straight line equations are indicated by the arrows. Inset depicts temperature variation of the percentage strain developed along the a axis.

Fig. 4. Variation of unit cell volume with temperature. The straight line is fit to the data points in the temperature range 80 K to 293 K. Inset shows the temperature variation of the excess volume (ΔV).

Fig. 1 Awana et al.

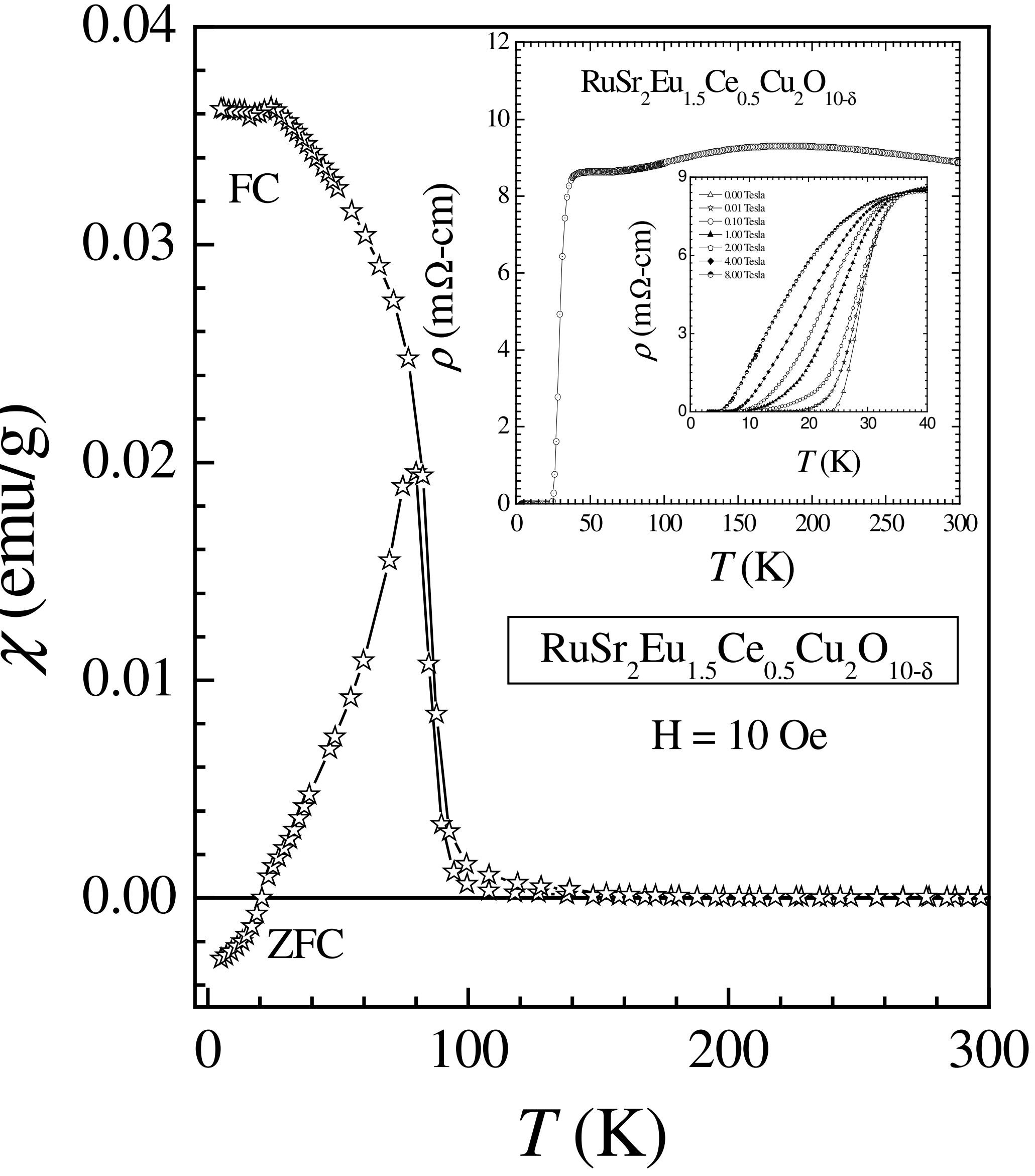

Fig. 2

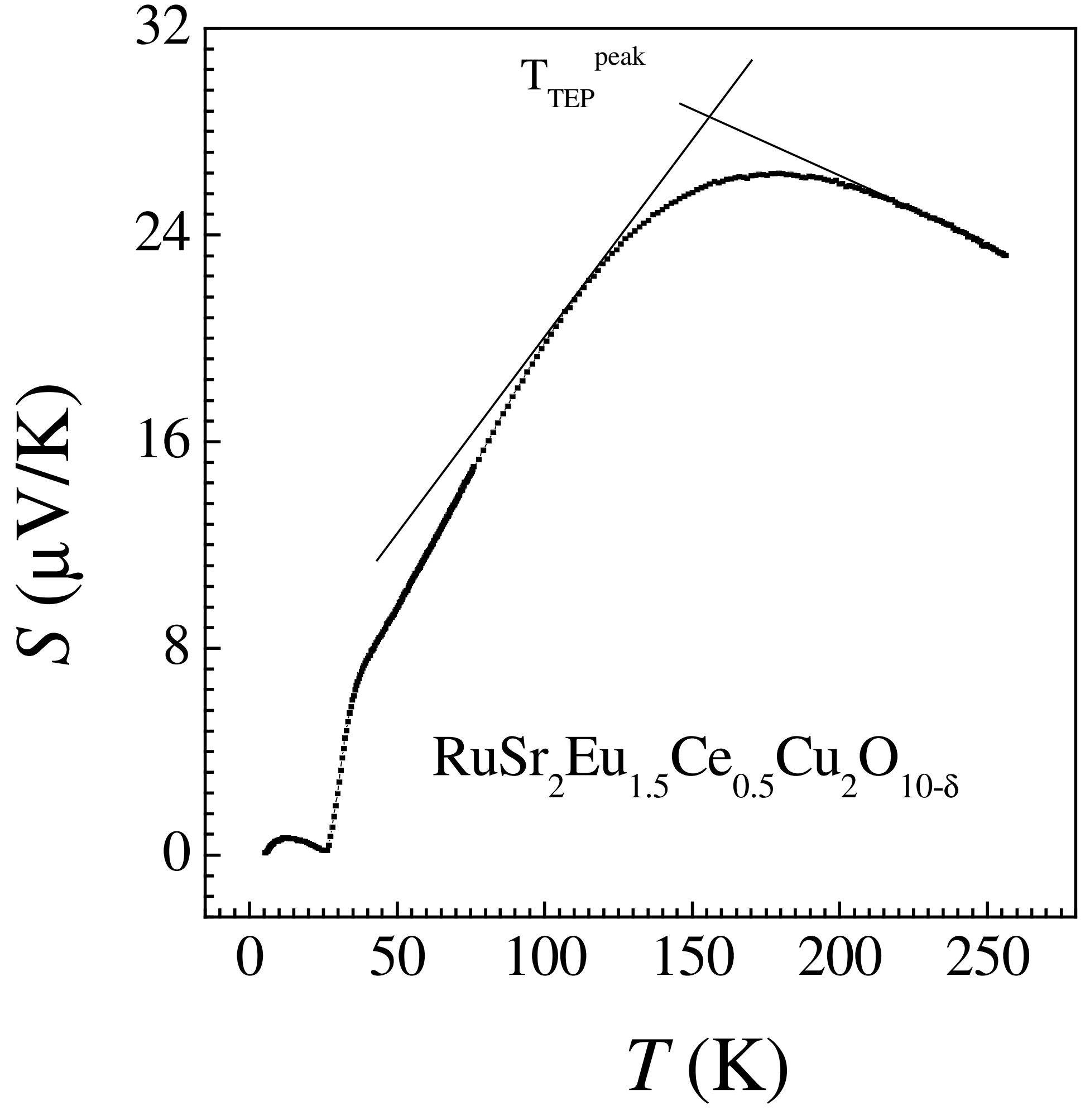

Fig. 3 Awana et al.

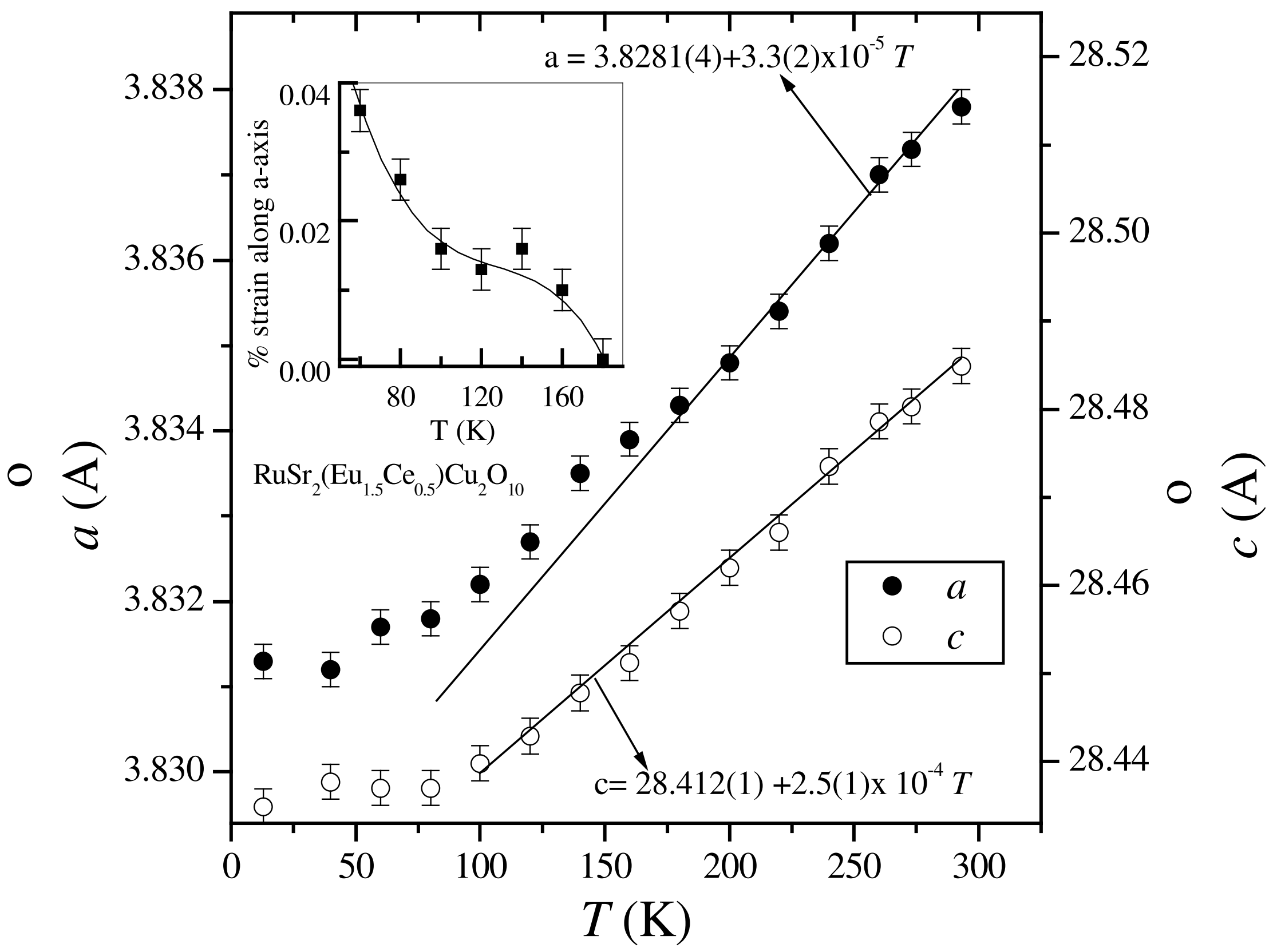

Fig. 4 Awana et al.

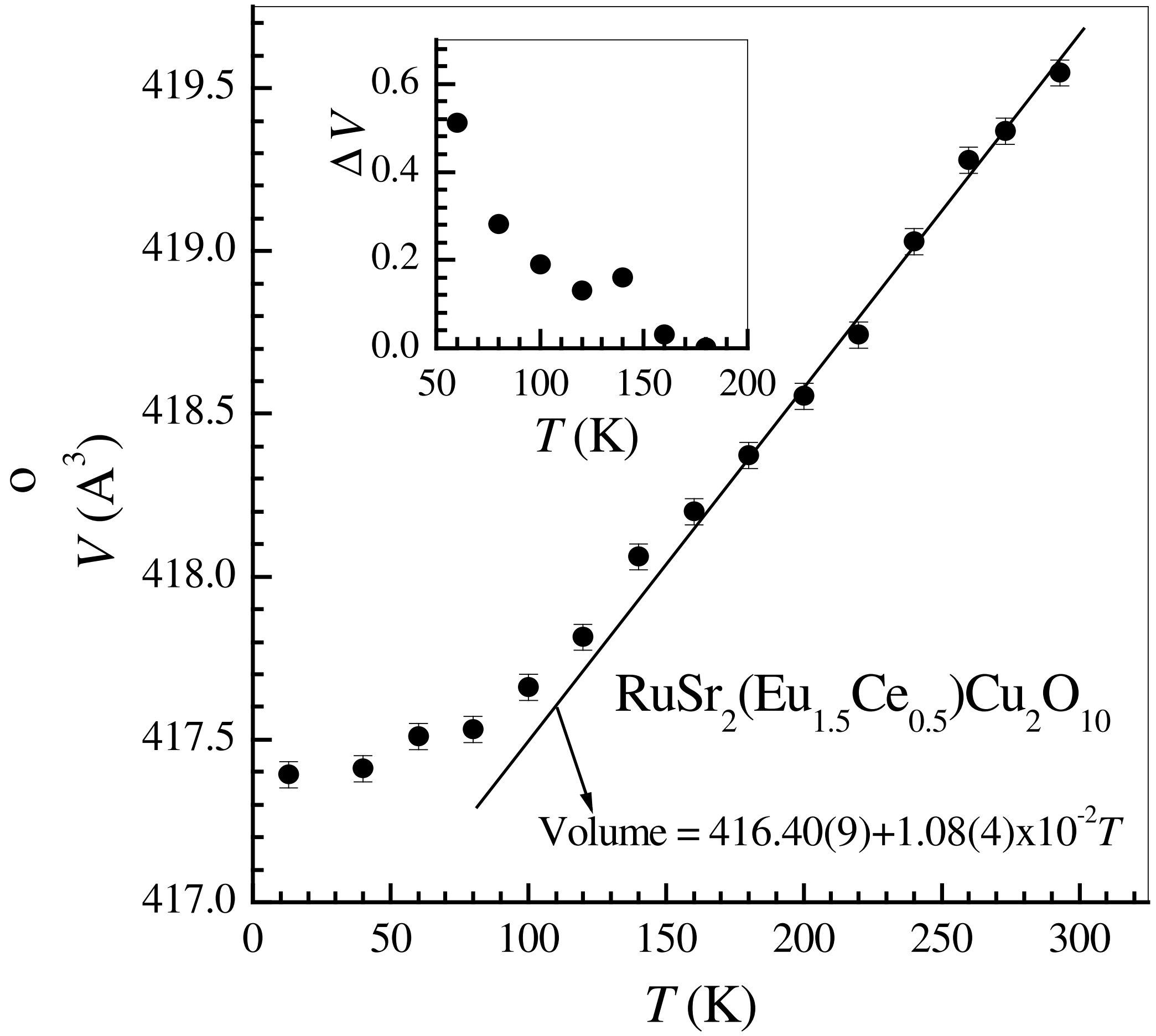